\documentstyle[pre,epsfig,aps]{revtex}
\title{Slow nucleic acid unzipping kinetics from sequence-defined barriers}
\author{S. Cocco$^1$, J.F. Marko$^{2}$, R. Monasson$^3$}
\address{
$^1$ CNRS--Laboratoire de Dynamique 
des Fluides Complexes, 3 rue de l'Universit{\'e}, 67000 Strasbourg, France;\\
$^2$ Department of Physics, 
The University of Illinois at Chicago, 845 West Taylor Street, Chicago, IL
60607-7059;\\
$^3$  CNRS--Laboratoire de Physique Th{\'e}orique de l'ENS, 24 rue
Lhomond, 75005 Paris, France.} 

\begin{document}
\maketitle

\abstract{
Recent experiments on unzipping of RNA helix-loop structures
by force have shown that $\approx 40$-base molecules can undergo kinetic 
transitions between two well-defined `open' and `closed' states, on a 
timescale $\approx 1$ sec [Liphardt {\em et al.}, 
Science 297, 733-737 (2001)].   Using a simple 
dynamical model, we show that these phenomena result
from the slow kinetics of crossing large free energy barriers 
which separate the open and closed conformations.
The dependence of barriers on sequence along 
the helix, and on the size of the loop(s) is analyzed. Some DNAs and RNAs 
sequences that could show dynamics on different time scales, 
or three(or more)-state unzipping, are proposed.}

\section{Introduction}

Helix-loops are the basic secondary-structure elements of folded
single-stranded nucleic acids (ssNA). Recent physical studies of
single helix-loop RNAs have revealed that despite their simple
structures, they can display interesting dynamics\cite{lip}.  When
placed under moderate tensions $\approx 15$ pN, telegraph-noise-like
`switching' behavior can be observed.  The characteristic time of this
switching has been seen to be on the $\approx 1$ sec timescale,
surprisingly large given the small size ($\approx 10$ nm) of the
molecules.  The purpose of this paper is to present a simple theory
capable of reproducing these slow switching kinetics.

The model we use for the energy of a helix-loop structure
considers states of partial `unzipping'\cite{Pol}. It
is based on available quantitative descriptions
of base-pairing interactions and single-strand nucleic acid
elastic response. In general a
large free energy barrier between open and closed states for
helix-loop structures is found, which qualitatively explains the
observed two-state switching kinetics. We then combine this
model with Eyring-Kramers transition-state theory to access the 
kinetics, and show how this barrier requires a long timescale 
to be crossed.  Finally, we discuss how essentially the
same approach can be used to describe branched-helix structures, and
we predict three-state-switching for a specific molecular
architecture.

\section{Model of Base-Pairing and ssNA Elasticity}

Our description uses the series of molecule configurations obtained by
successively breaking base-pairs, starting from the molecule 
ends\cite{Pol}.  In addition to the base-pairing free energy of
the double-stranded part of the molecule, we take into account the
elastic response of the extended, unpaired, single-stranded part of
the molecule\cite{Ess,Noi,Hwa,Lub}. The free energy change associated
with opening base pair $i$ is simply the difference between the free
energies of the paired region, and the free energy associated with
extension of the open region,
\begin{equation}
\label{dg}
\Delta g(i,f)=g_0(i)-2\,g_s(f,i)
\end{equation}
Here $g_0(i)$ is the free energy of opening base pair $i$
(between $1$ k$_B$T and 5 k$_B$T), obtained
using the MFOLD program \cite{Zuk} .
The stretching contribution, $g_s(f,i)$, is the free energy per base
at constant force for stretched ssDNA, which provides a reasonable
estimate for ssRNA.   Integration of an empirical fit to the
experimental ssDNA force-extension\cite{Smi} gives:
$g_{\rm s}^{FJCL} (f) = 
k_B T {l_{ss}}/{d}\,\ln [\sinh(u)/u]$
with $u\equiv d \, f / [k_B T]$ and 
parameters $l_{ss}\simeq 5.6$ {\AA}, $d=15$ {\AA}.
Internal unpaired bases (Fig.~1) are considered to open along
with the base pair immediately preceding them; those base opening 
steps therefore pick up a multiple-base $g_s$ contribution.
The free energy to unpair the first $n$ base pairs is just the
sum of the free energies of the individual base-opening steps,
\begin{equation}
G(n,f)=\sum_{i=1}^n \Delta g (i,f)
\label{G}
\end{equation}
We emphasize that both the base-pairing and the elasticity contributions
to (\ref{dg}) and (\ref{G}) are free energies, i.e. are coarse-grained
over atomic-scale fluctuations. The above zipper model \cite{Pol} is
roughly equivalent to that described in the Supplementary Materials
of \cite{lip}; the presence of the stretching force allows to
discard complex opening/closing pathways 
relevant at zero force\cite{Isam}.  Note that some cooperativity 
between base pairs is introduced by MFOLD (the pairing free energy
of neighboring base pairs is not simply additive, but includes some
sequence specific stacking interactions).

\section{Equilibrium Behavior of a Single Helix-Loop Structure}

Fig. 2a shows $G(n,f)$ and the equilibrium probability  
$P(n,f) \propto \exp(- G(n,f)/k_BT)$ for a simple RNA hairpin, 
called P5ab\cite{Cate} (Fig. 1a), under the solution
conditions of \cite{lip} near the critical force $f^*\approx 15$ pN
where two-state switching is observed.  At the critical force, the
open and closed states ($n=22$ and $n=3$, note $n$ is just the number
of broken base pairs) dominate; below or above this
force, the free energy landscape is tilted either to favor the $n=3$
or $n=22$ state. At the critical force, there are various barriers 
due to drops in free energy resulting from openings of the U bulge, 
the weak non Watson-Crick GA central pairs, and the final loop.
The largest barrier $\simeq 11$ k$_B$T must be crossed to reach
the closed state from the open one, and vice-versa.
The two-state behavior observed for P5ab follows from the
partition of probability into two peaks separated
by an improbably-accessed barrier region (see Supplementary Materials,
Ref. \cite{lip}).

\section{Dynamical Model}

To reach a quantitative understanding of switching, we introduce
a dynamical model for the motion of the `fork' separating 
the base-paired and opened regions of the molecule\cite{Kra}.
We propose the following expressions for the 
elementary rates of opening and closing base pair $n$
({\em i.e.} to move the boundary between the open and closed 
portion of the molecule from $n$ to $n-1$ or to $n+1$): 
\begin{equation}
\label{rate}
r_o (n)= r\; e^{-g_0(n)/k_B T} \quad , \qquad
r_c (f,n)=r \;e^{-2 g_s(f,n)/k_B T} 
\end{equation}
Here $r$ is essentially the microscopic rate for a base pair
to move together or apart in the absence of 
tension or base-pairing interactions, or roughly 
the inverse self-diffusion time for a few-nm-diameter object\cite{Doi}, 
$r \eta \ell^3 / k_B T \approx 10^7$ s$^{-1}$, with
$\ell = 10$ nm, $\eta=0.001$ Pa sec, and $k_B T = 4\times 10^{-21}$~J.
 
In (\ref{rate}) we have made the simplifying approximation that
the opening rate $r_o$ has no force dependence, and is simply
proportional to the exponential of the base-pairing free energy of
(\ref{dg}).  
Eyring--Kramers transition--state theory applied to breaking of a
chemical bond considers indeed the potential energy to be `tilted' by
a force-times-displacement contribution. 
Because hydrogen bonds break for relatively small displacements ($\approx
0.1$ nm) the reduction in the potential energy of the
single-base-opening transition state will be roughly 
15 pN $\times 0.1$ nm = 0.3 $k_B T$. This can be neglected with respect 
to the base-pairing free energy of a few $k_BT$, which is a lower 
bound to the energy of the transition state
associated with breaking a single base pair. 
Detailed balance then determines the closing rate
$r_c$ to be proportional to the exponential of force times
displacement, {\em i.e.} to the energy of a fluctuation that is able
to pull the two bases back together in opposition to the applied force.

The rates (\ref{rate}) lead to a master equation for the probability
$\rho _n (t)$ for the boundary to be at site $n$ at time~$t$:
\begin{equation}
\label{master}
\frac{d\, \rho _n(t)}{d\,t}=-\sum_{m=0}^{N}\; T_{n,m}\; \rho_m (t)
\end{equation}
This $(N+1) \times (N+1)$  matrix $T_{n,m}$ is tridiagonal, with nonzero 
entries $T_{m-1,m}=r_c(f,m)$, $T_{m+1,m}=r_o(m)$, and 
$T_{m,m}=-T_{m-1,m}-T_{m+1,m}$.

\section{Switching Kinetics of a Single Helix-Loop Structure}

We have solved (diagonalized) (\ref{master})
for P5ab (Fig. 1a). The smallest eigenvalue is 0; the
eigenvector is the equilibrium Boltzmann distribution.   
At the critical force
where the molecule is on average half-open, the smallest non zero 
eigenvalue is $\lambda _1=
2.1 \times 10^{-6} r$, which corresponds to the slowest 
mode of fluctuation, the `switching' of the boundary of the 
open region between $n \approx 3$ and $n\approx 22$.   
The remaining 21 eigenvalues are all well separated from the leading ones.
The second largest eigenvalue is $\lambda_2=0.9 \times 10^{-4} r$ (Table).
Thus the theoretical dynamics of P5ab involve one slow opening-closing 
transition, combined with many other transitions occurring more than 50 
times faster.
The net rates of opening
($k_o$) and closing ($k_c$) can be computed from 
$\lambda _1= (k_o+k_c)$, and the ratio of the
open and closed equilibrium probabilities equal to $k_c/k_o$.
To compare our theoretical results with the experiments of \cite{lip} we have
fitted our result for $k^*\equiv k_o=k_c$ at the force $f^*$ where
the open and closed states have equal probability
to experimental data\cite{noter}, giving $r = 3.6 \times 10^6$ sec$^{-1}$.  

Fig.~3a shows time series from Monte Carlo simulations of the molecular 
motion; slow two-state
switching is seen on a $\approx 0.25$ sec timescale, 
on top of which occur much faster small fluctuations.
When we convolve these data with a 20 Hz low-pass filter 
(as used experimentally\cite{lip}) the result (Fig. 3b) is
essentially the same as the experiment.
The variation of the rates with force given by theory
are also in good agreement with experiment(Fig. 4). 
The `transition state' coordinate can be inferred from the
relative slopes of the opening and closing rates of Fig. 4 
(independent of the fitted parameter $r$) around the
critical force\cite{lip}.  The critical-force transition state 
is at the top of the free energy barrier,
at $n^*=11-12$ (Fig. 2a).

\section{Barriers from generic helix-loop structure}

For molecules with repeated sequences and no terminal loop e.g.
AU followed by a long GC helix or a crosslink between end bases, 
the kinetics essentially corresponds to a diffusion in a 
flat free energy landscape, and 
shows none of the two-state character of the P5ab RNA.  
For a repeated AU sequence of $n=25$ base pairs, 
the switching time is just the diffusion-like time 
$t^* \approx 2 N^2/(\pi^2 r_o^*) = 2\times 10^{-4}$ sec (Table).  
Thus, the 1000-times longer switching time of P5ab comes from the 
$11$ k$_B$T barrier of Fig. 2a. 

The presence of a loop is sufficient
to generate such a large barrier and, consequently, two-state switching.
The simplest illustration is given by a homogenous DNA sequence ending 
with a loop. We have considered a 24-base Poly(GC) homogeneous-sequence helix 
terminated with a 4-base loop (Fig. 1b, 2b)\cite{Bus}. Our theory predicts 
a two-state switching behavior on the same time scale
as P5ab.  The switching time is 50 times larger for a longer 8-base loop
(Table). 

The free energy barrier $G^*$ at criticality for a 
S-base-pair stem (uniform pairing free energy $g_0$) followed by a L-base
loop (closing free energy $g_{loop}(L)$ at zero force) (Fig.~1) can
be simply estimated.  The critical force $f^*$
is given by the condition that the free energy of the open molecule
equals the free energy  of the closed molecule,
$G(0,f^*)=G(S,f^*)=S\, g_0 - (2S+L)\,g_{ss} (f^*)- g_{loop} (L)$.
The barrier height $G^* (S,L)\equiv G(S-1,f^*)$ then reads
\begin{equation}
G^*(S,L)=  (S-1) (g_0-2 \,g_{ss} (f^*) )=\frac{(S-1) 
\left( g_0\; L +2 \;g_{loop}(L) \right)}
{L+2 S} \qquad . \label{cori}
\end{equation}
Table~1 shows that, for
a fixed stem length $S$, the critical force decreases with the length
$L$ of the loop, while the free energy barrier $G^*$, and the switching
time $t^*$ increase.    
For non-repeated sequences, (\ref{cori}) is only approximate
when substituting $g_0$ with an average 
pairing free energy; it allows an estimate of how the switching
time depends on $S$ and $L$ \cite{noterandom}. Note that
the critical barrier essentially depends on the smaller of the two
lengths $S,L$. 

\section{Branched-Helix Molecules and Multiple-State-Switching}

Our approach can be extended to more complicated situations
e.g. nucleic acids with branched structure.
An example is the P5abc$\Delta$A RNA molecule\cite{lip} of Fig. 1c, 
with free energy landscape as shown in Fig. 2c.  
The opening of the
first 12 bases (helix H1) follows as above, but going past the branch, 
description of the independent opening of the two helix regions
requires a three-dimensional free energy landscape
{\em i.e.} free energy as a function of the positions of the two
unzipping boundaries\cite{notep5abcDa}.  
A rich behavior emerges, shown in Fig.~3c.
At the critical force $f^*=12.9$~pN, H1 switches on a 
long time scale $t^*=10$~sec (Table), while the short lateral helix (H2)
opens and closes with a much shorter characteristic time 
$t_2 \simeq 9$~msec.  The lateral long helix (H3) 
opens very rarely at this force. Predictions for the rates, barrier height, ...
are reported in Table~1.

It is possible to design molecules with multiple-state
dynamics.  Consider the molecule of Fig. 1d,
which has two well-bound Poly(GC) regions separated by an unpaired
bubble, and terminated by a loop.  For this system, three-state
switching occurs (Fig. 3d),
and should be observable in the frequency range accessible
in experiments such as that of \cite{lip}.

\section{Conclusion}
A few improvements might be added to this model.
First, opening may occur through the cooperative
nucleation of a few base pairs bubble\cite{Noi}. 
Second, mismatches might take place
during closing \cite{Ansari}, although they are highly limited
by the presence of the 15 pN force.
Finally it would be interesting to be able to describe unzipping events
involving breaking of tertiary structures; these are thought to be
present in the molecule P5abc$\Delta$A in presence 
of Mg$^{2+}$ \cite{lip,Fan}.

A general result of our work is that slow switching
character should be quite generic for small biological
molecules with helix loop structure.
We note that our approach could also be used to analyze the 
opening-closing dynamics of nucleic-acid-detecting DNA 
`beacons'\cite{Tya}, both on their own, and in the presence 
of their targets. In this case the `unzipping' forces are 
applied by the hybridization interactions instead of by a 
large force transducer.
Since such experiments amount to molecule recognition 
processes it is not impossible that slow barrier-crossing
transitions of the sort discussed here occur in-vivo.

\section{Acknowledgements}
We thank P. Cluzel for helpful discussions and M. Zuker for advice
on use of MFOLD.  Work at UIC was supported by NSF Grants
DMR-9734178 and DMR-0203963,
by the Petroleum Research Foundation of the American
Chemical Society, by the Research Corporation,by a Biomedical
Engineering Research Grant from the Whitaker Foundation, and by a
Focused Giving Grant from Johnson \& Johnson.  S.C. is partly funded
by an A. della Riccia grant.  R.M. is supported in part by the
MRSEC Program of the NSF (DMR-9808595). 

\begin{table}

\begin{center}
\begin{tabular}{|c|c |c| c|c|c|c|c|c|}
\hline
Molecule &  $\Delta G$ & 
$f^*$ & $\ln ( k^{*}$ & 
$t^{*}$ & $\ln  k_{o} (f)$ &
$\ln  k_{c} (f)$ & $G^*$ & $\lambda _1 /\lambda _2$ \\ (N)
& k$_B$T & (pN) &/r) & (sec) & (sec$^{-1}$) &
(sec$^{-1}$) & (k$_B$T) & \\
\hline \hline
{\hbox{\rm P5ab}} &   57.2 & 15.1 & -13.8 & 0.25 & 
-42.9+1.93 f & 27.5 -2.74 f & 10.7  & 0.023 \\
(49)&  {\bf 66.7}  & {\bf 14.5}  &   &{\bf 0.25} & {\bf -39$\pm$2.3 $-$}  
 & {\bf 41$\pm$1.9 $+$}  & & \\   & {\bf $\pm$8.5} & {\bf $\pm$1} 
& & {\bf $\pm$0.1} & {\bf (2.9$\pm$0.2)f} & 
{\bf (2.8$\pm$0.1)f }& & \\ 
\hline \hline
{\hbox{\rm Repeated}} &&&&&&&& \\
{\hbox{\rm AU}} (50) & 44.8  & 12.3  & -6.7 & 0.0002 & & &0 & 0.99\\
{\hbox{\rm GC}} (50)  & 135 & 27.4  & -10.3 & 0.008 &&& 0 & 0.99 \\
\hline \hline
{\hbox{\rm Poly(GC)}} &&&&&&&& \\
{\hbox{\rm no loop}} (24) & 41.3 & 19.7 & -6.5  & 0.0002 &&& 2.4 & 0.42 \\
{\hbox{\rm 4b loop}} (28) & 34.8 & 15.6  & -14.1 & 0.37 & -46.7 + 2.02 f&
1.7 - 1.01  f& 11.3 & 2  $10^{-4}$\\
{\hbox{\rm }}   & {\bf 43$\pm$3} & {\bf 16}  &  &&&& & \\
{\hbox{\rm 8b loop}} (32) & 33.2 & 13.7 & -18.0 & 16.7 & -46.8 + 2.1 f&
1.3 -1.4  f& 15.7& 3 $10^{-6}$ \\
{\hbox{\rm double}}  (56) & 71.3 & 15.8  &-14.4  & 0.48 & -53.2+2.26 f
& 8.34-1.44 f& 10.9 & 0.38 $^{a}$
\\ \hline \hline
P5abc$\Delta$A & 70.6  & 12.9  & -17.1 & $10^{b}$ & 
-43.8 + 2.06  f & 9.4 -2.05  f& 14.3  & 9 $10^{-4}$\\
(64) & {\bf 71.9}  & {\bf 12.7 } &  & & 
{\bf -39$\pm$9.3 $+$} & {\bf 58$\pm$7.5 $-$} && \\
& {\bf $\pm$11.5 }  & {\bf $\pm$0.3} &  & &
{\bf (2.7$\pm$0.7)f} & {\bf (4.2$\pm$0.5)f} && \\
\hline\hline 
\end{tabular}\end{center}

\caption{Theoretical results for the molecules of Fig. 1 compared to
experimental values from [1, 14] (in bold) when available. Columns
indicate the free energy $\Delta G=G(N,f=0)$ at zero force, the
critical force $f^*$, the rate of opening/closing $k^*$, the switching
time $t^*$ and the free energy $G^*$ of the highest barrier at
criticality, the variation of the opening and closing rates upon force, 
and the ratio of the two largest non zero
eigenvalues. This ratio is small when open and closed states are well
defined, and close to one otherwise.  Uncertainties in base-pairing
free energy are at most $\delta g\approx 0.5 k_B T/$bp, with
consequent total uncertainties of $\approx N^{1/2}\delta g$ (or about
$3 k_B T$ for $N=25$) for $\Delta G$ and of $\approx (N/2)^{1/2}\delta
g \simeq 2 k_B T$ for $G^*$, and $\approx 3 k_B T/20$ nm$\approx 0.6$
pN for the critical force.  Notes: $^a$ the ratio is close to one due
to the presence of three, and not two, states giving rise to two large
barrier crossing times (the remaining fluctuation times are much
shorter: $\lambda _1/ \lambda _3 \simeq 0.0001$); $^b$ the predicted
rate (0.1/sec) for P5abc$\Delta$A is very close to the lowest
frequency (0.05 Hz) resolved experimentally [1], thus its value
is known only roughly.  }
\end{table}

\begin{figure}
\begin{center}
(1a) \includegraphics[height=200pt,angle=0] {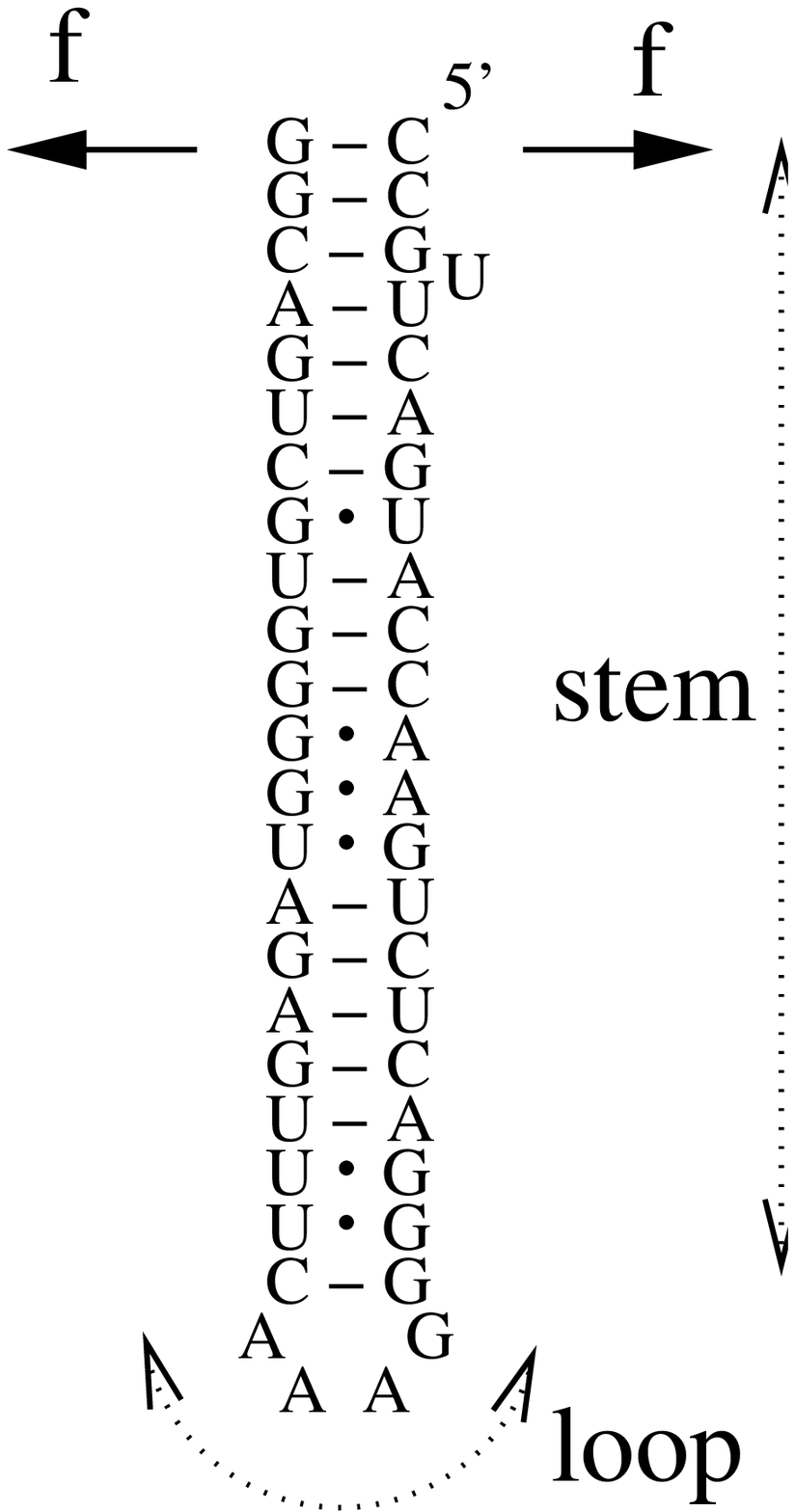}
\hskip 3cm
(1b)\includegraphics[height=180pt,angle=0] {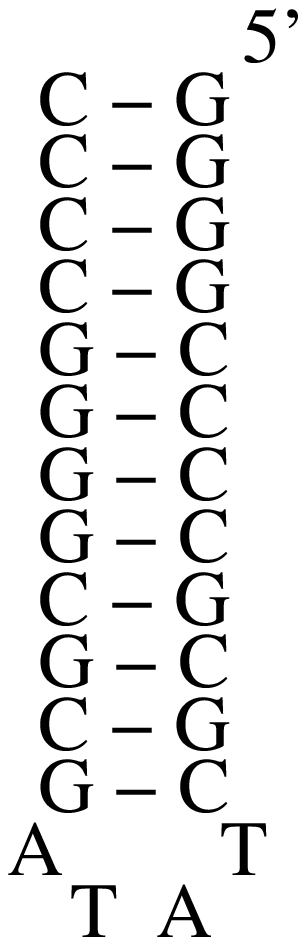}
\vskip 1cm
(1c) \includegraphics[height=200pt,angle=0] {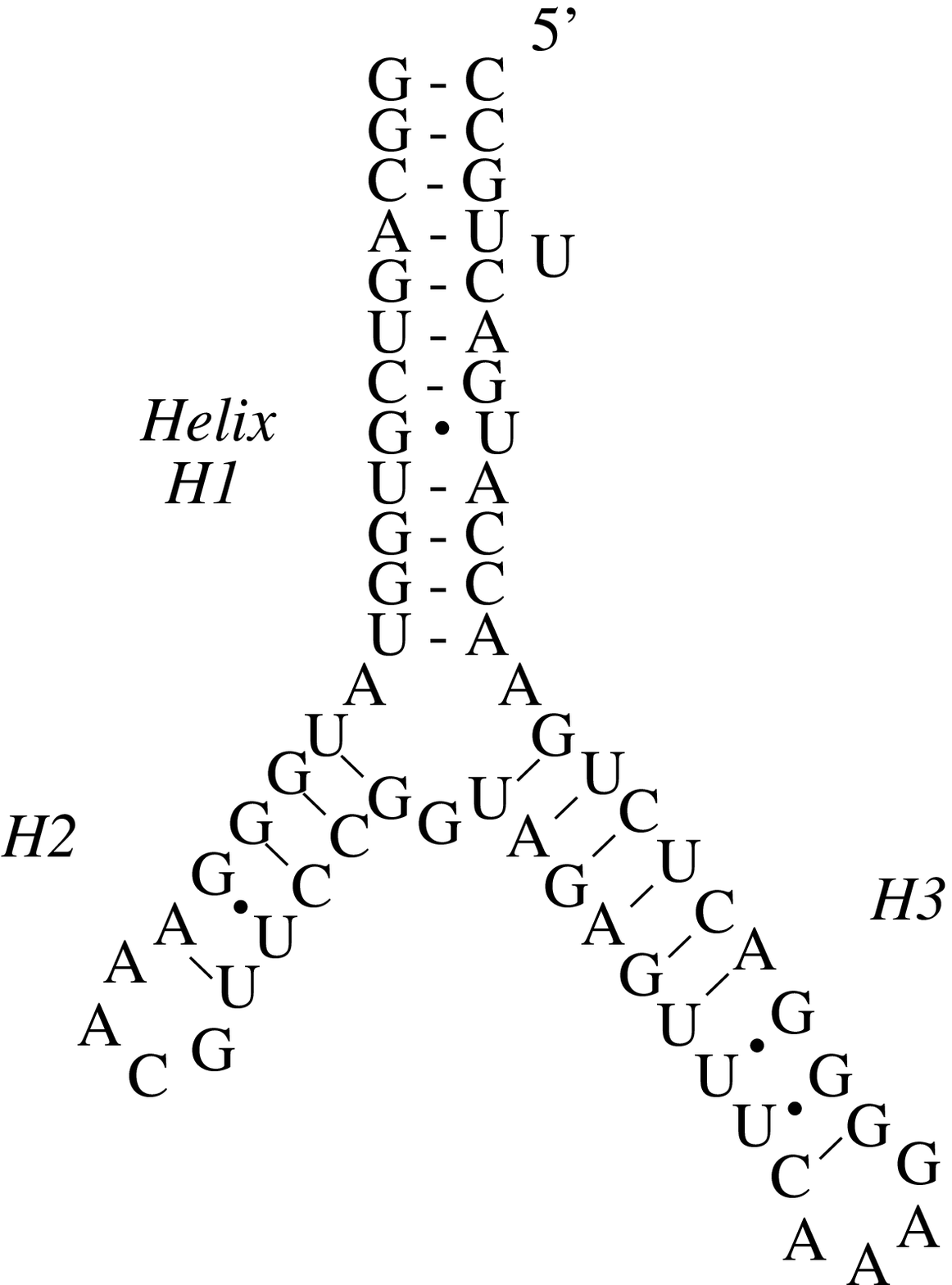}
\hskip 3cm 
(1d) \includegraphics[height=200pt,angle=0] {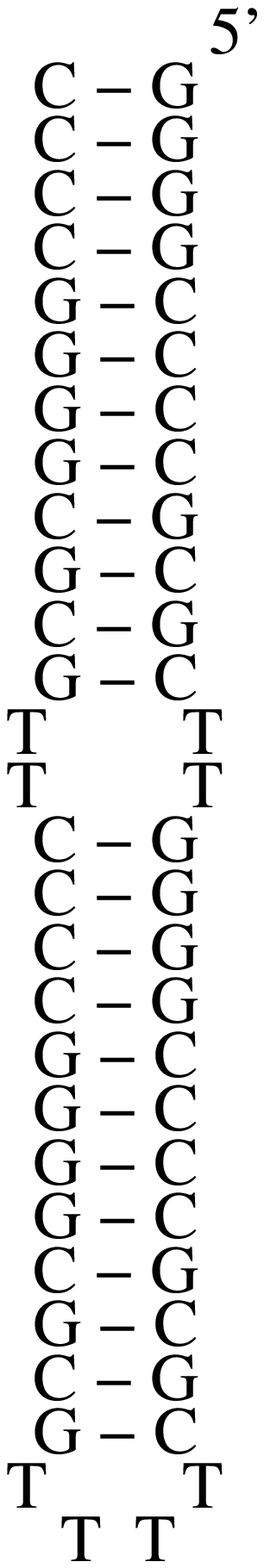}
\end{center}
\caption{
Nucleic acid unzipping experiment and molecules studied. 
A constant force is applied to a helix-loop
structure while the distance between molecule ends is measured.
(a) P5ab RNA,
a single helix-loop structure  present on the P4-P6 domain 
of a self splicing group I intron of the {\em Tetrahymena
thermophila}. The structure shown is predicted by Mfold [7] apart from
the G-A weak pairs [10] indicated here with dots, 
and with the U-bulge translocated; 
(b) DNA hairpin consisting of a poly(GC) helix terminated with an 
ATAT loop [14];
(c) P5abc$\Delta$A RNA, a variant of P5ab with an additional
helix giving a Y-branched structure at zero force;
(d) Hypothetical RNA molecule obtained by ligation of two poly(GC) helices
as in (b) and replacing A bases with Ts.
}
\label{figure1}
\end{figure}

\begin{figure}
\begin{center}
(2a) \includegraphics[height=200pt,angle=-90] {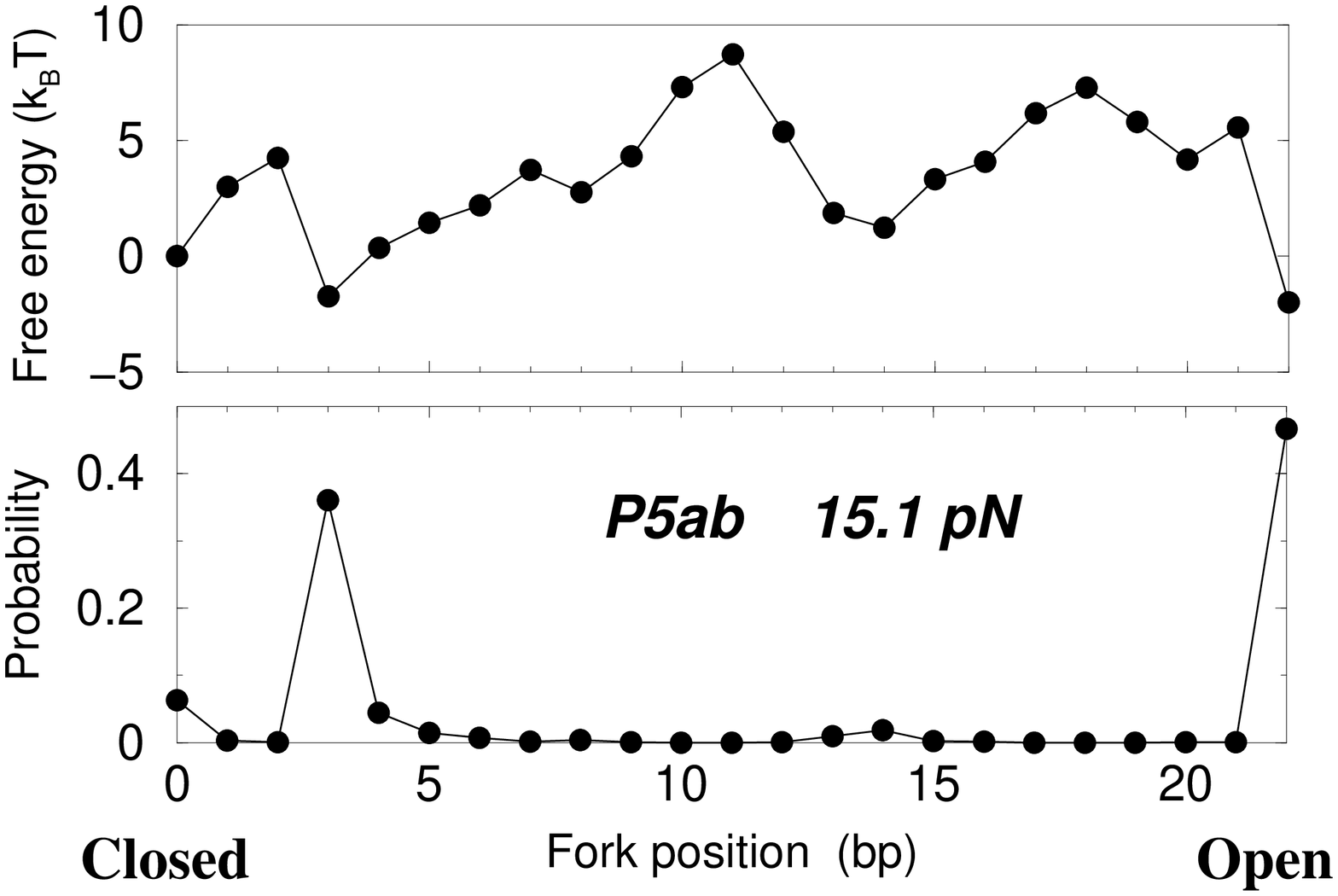}
\hskip .5cm
(2b) \includegraphics[height=200pt,angle=-90] {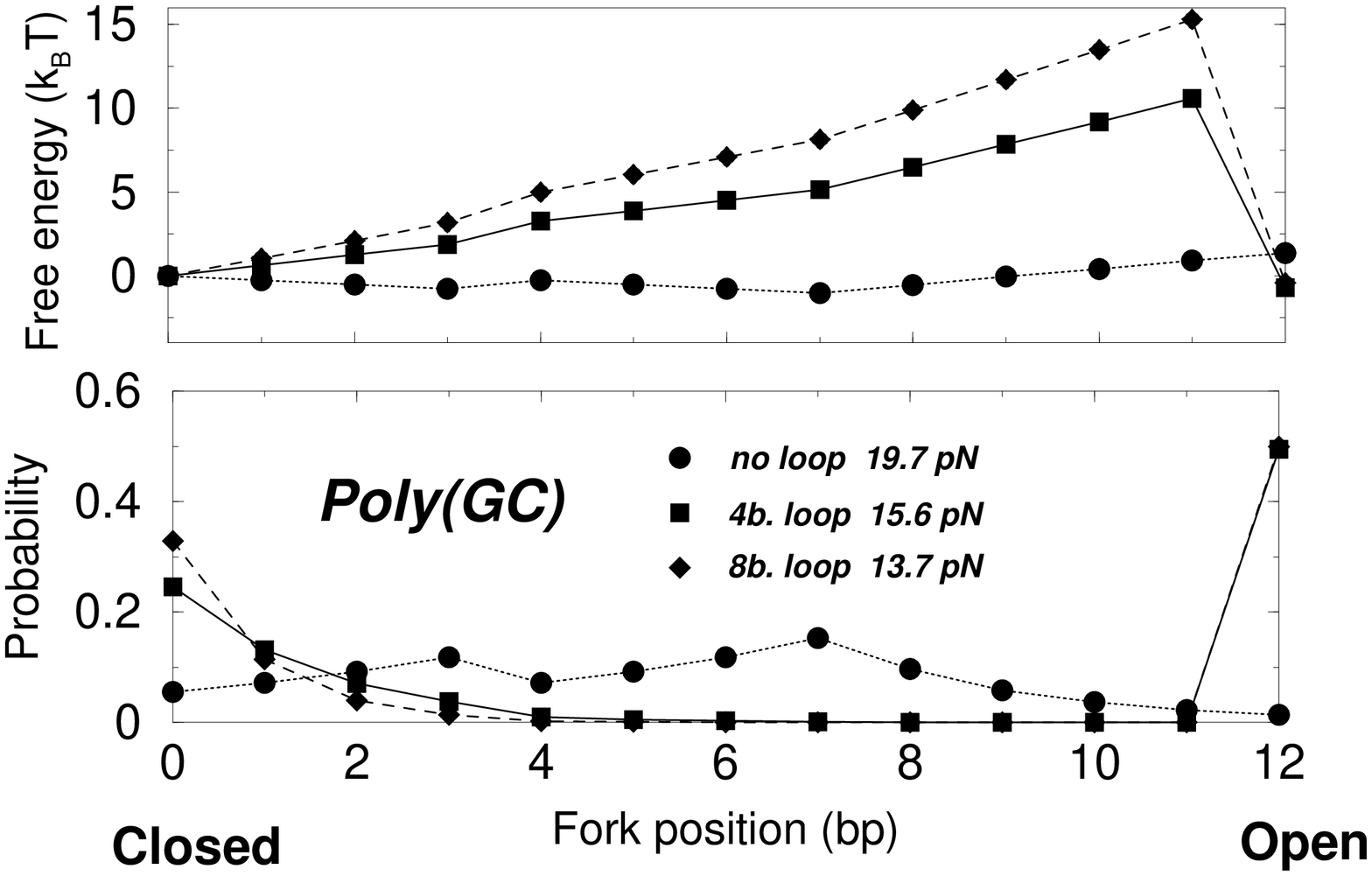}
\end{center}
\vskip 1.5cm
\begin{center}
(2c) \includegraphics[height=220pt,angle=-90]{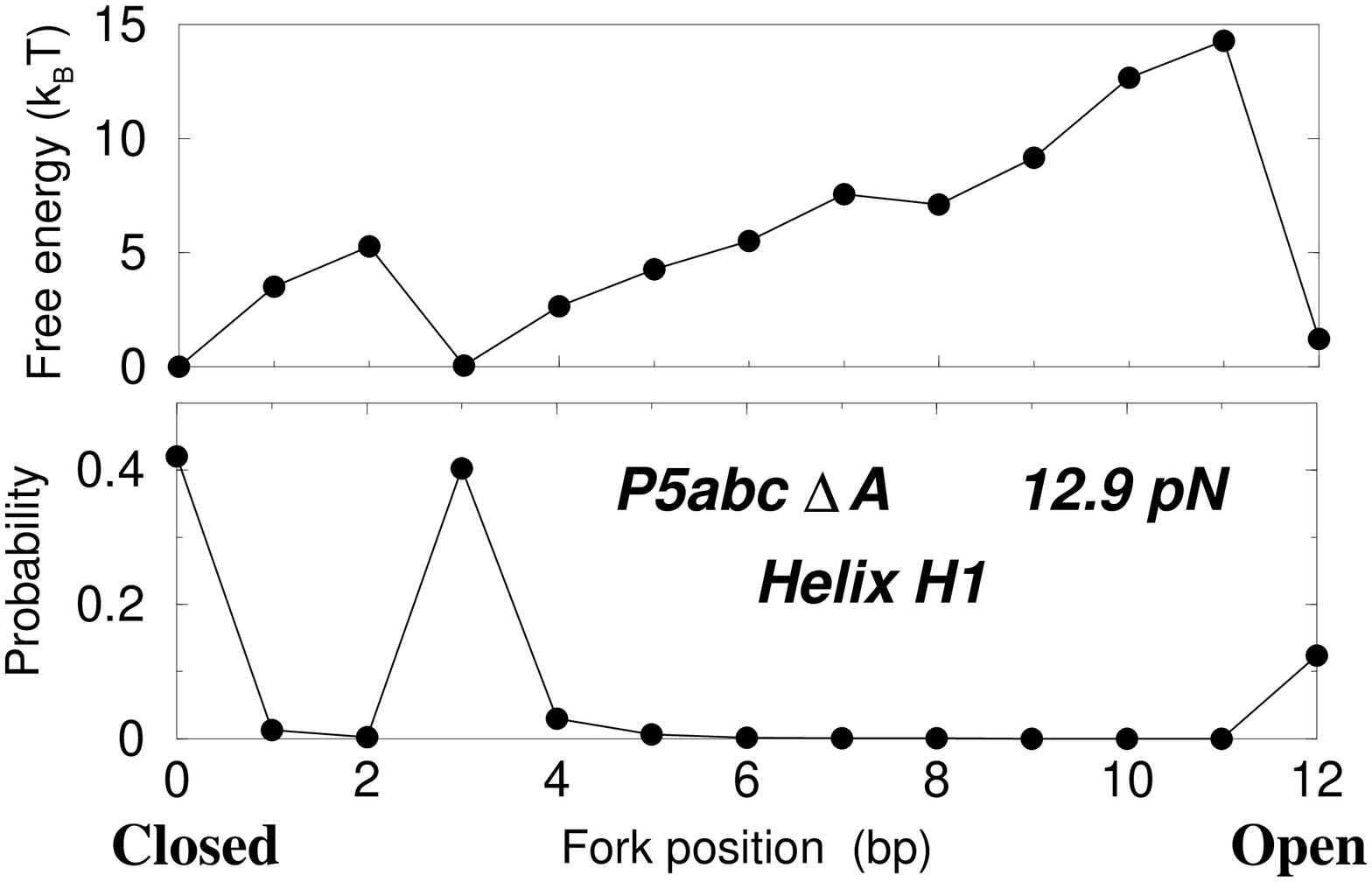}
\hskip 1cm
\includegraphics[height=170pt,angle=-90] {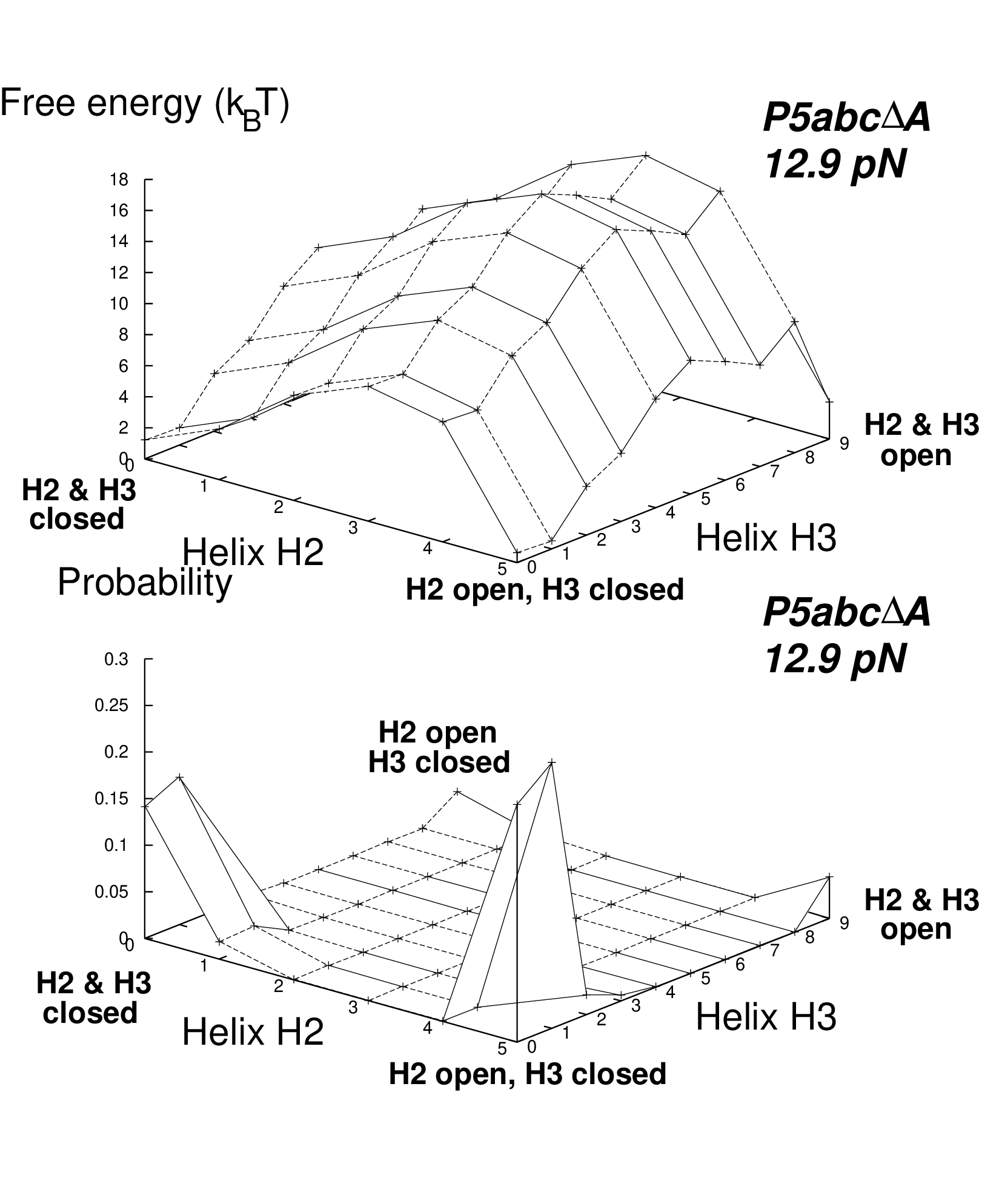}
\end{center}
\caption{Free energy landscapes (in k$_B$T, top) 
and probability distributions for the position of the opening fork 
(bottom) at the critical force for molecules of Fig. 1.  
(a) P5ab; (b) Poly(GC) RNA with: a 8bp loop (diamond), a 4 bp loop 
(squares); no loop (circles). 
(c) P5abc$\Delta$A; (left: opening
positions $n_1=0,\ldots , 12$ corresponding to helix H1, right:
opening positions $(n_2,n_3)$ with $n_2=0,5$ -- H2 -- and
$n_3=0,9$ -- H3 --); by definition configurations $n_1=12$ and 
$(n_2=0,n_3=0)$ coincide.}
\label{figure3}
\end{figure}
\newpage
\begin{figure}
\begin{center}
(3a) \includegraphics[height=200pt,angle=-90]{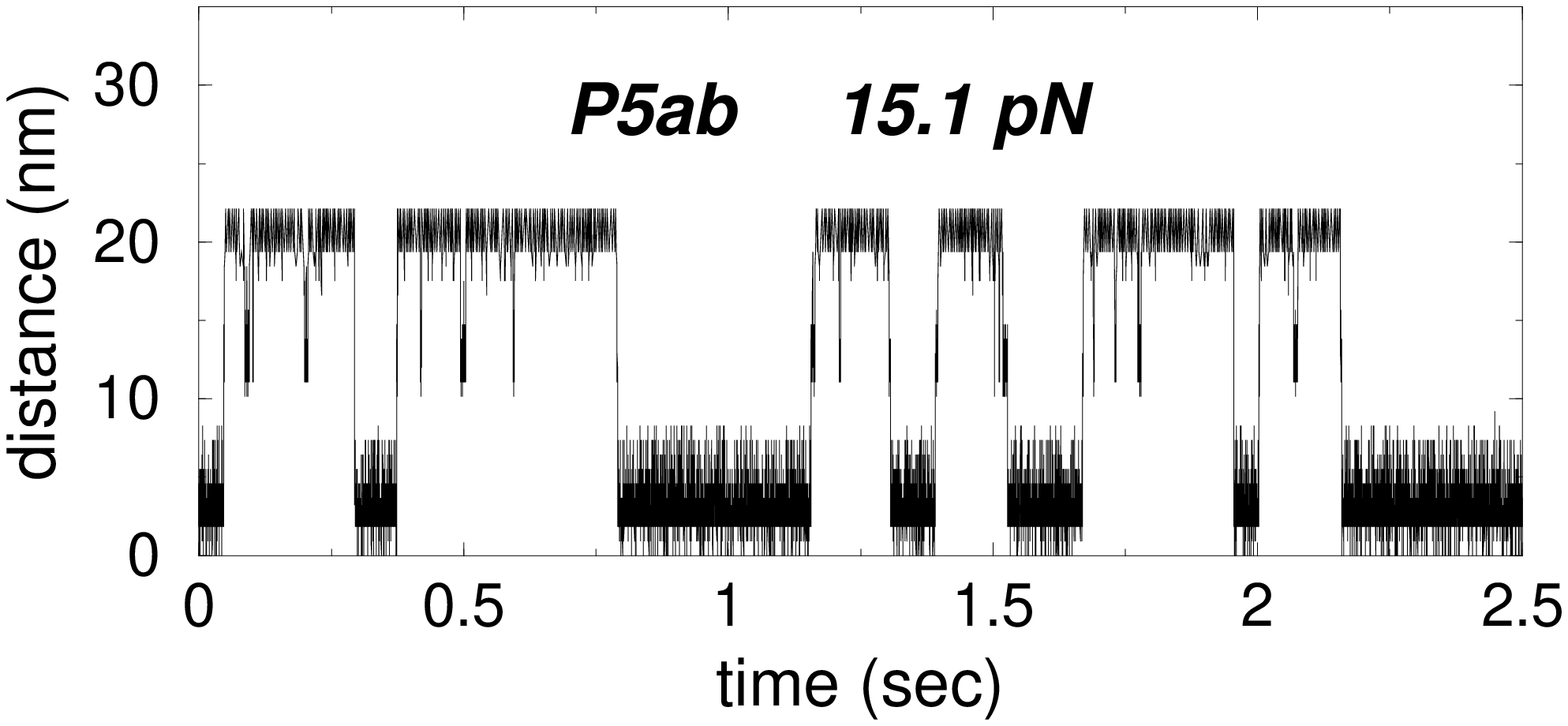}
\hskip .5cm
(3b) \includegraphics[height=200pt,angle=-90]{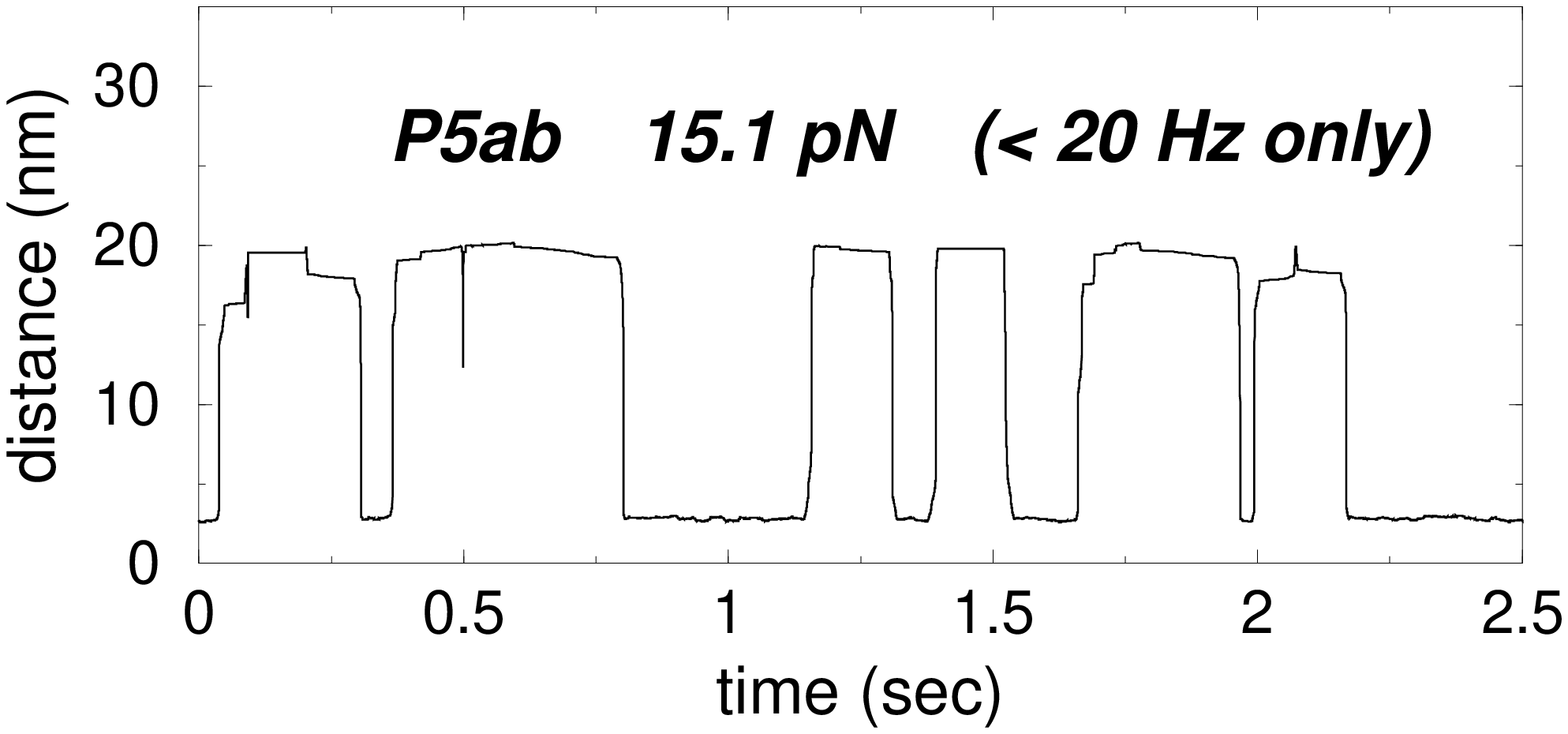}
\end{center}
\vskip .5cm
\begin{center}
(3c) \includegraphics[height=200pt,angle=-90]{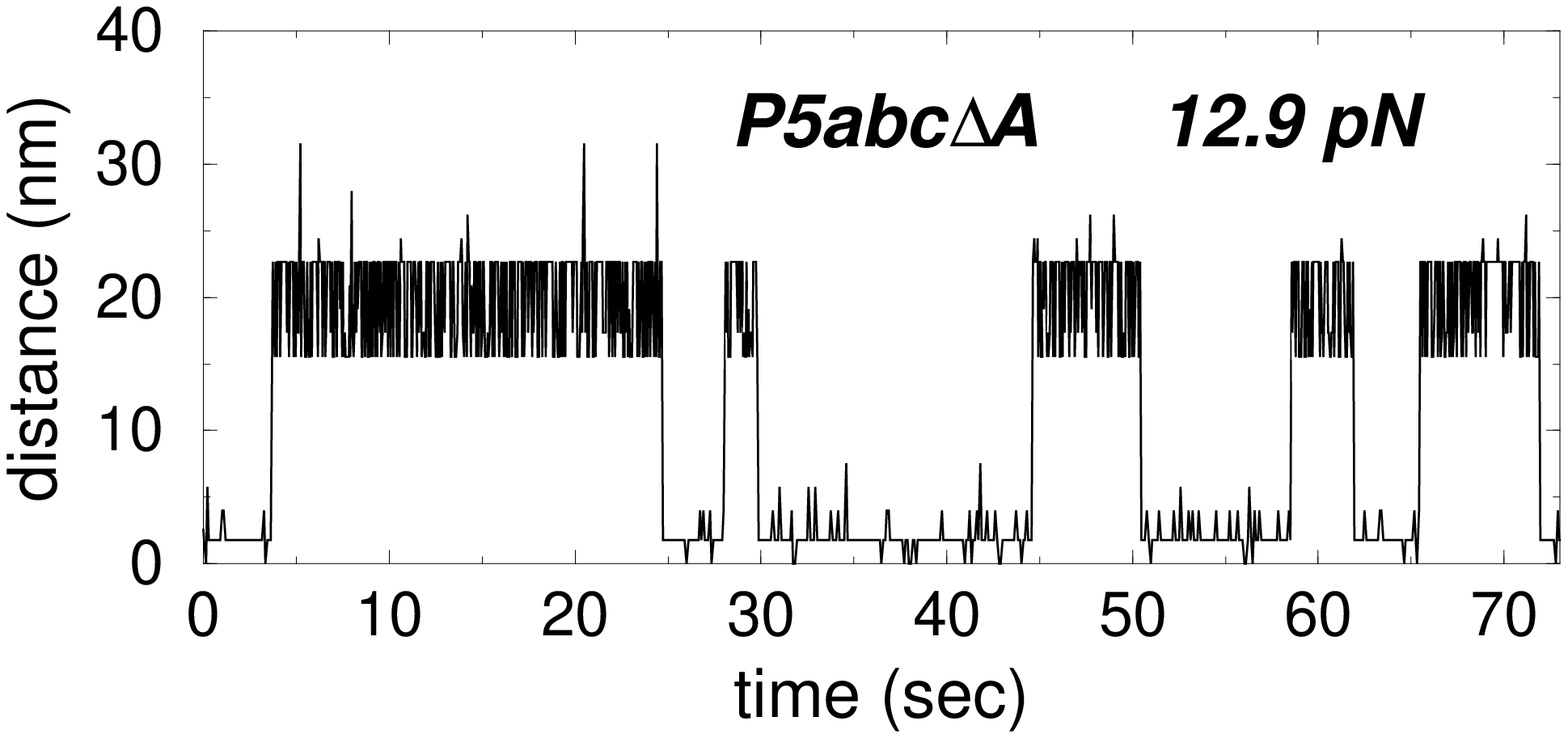}
\hskip .5cm
(3d) \includegraphics[height=200pt,angle=-90]{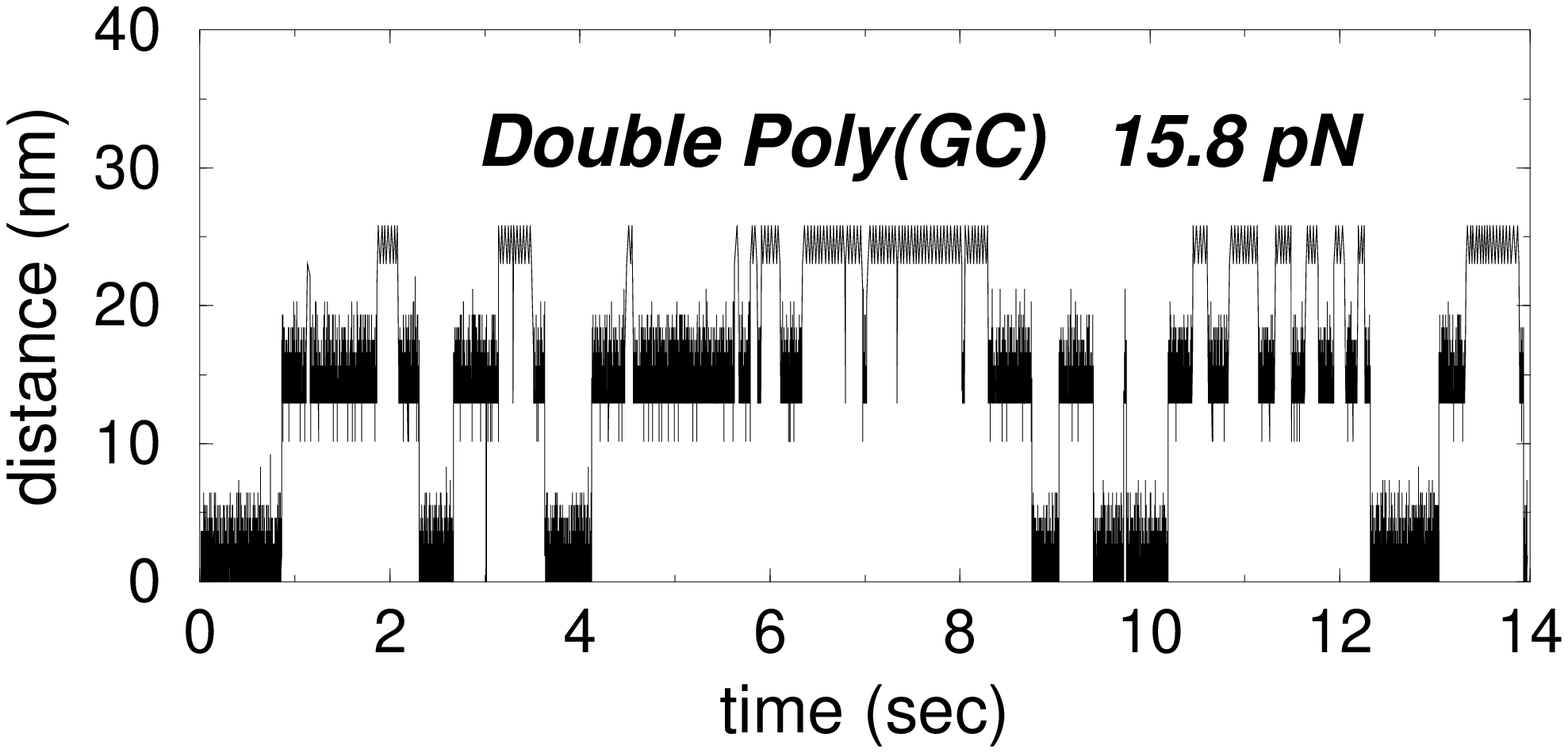}
\end{center}
\vskip .5cm
\caption{Unzipping kinetics at the critical force. 
Distances between extremities of the molecule 
are shown for: (a) P5ab; there is a slow switching 
between the
$n\simeq 3$ and $n \simeq 22$ configurations and fast transitions
between configurations $n$ around these ones (Fig 2a).
(b) Convolution of (a) where oscillations
faster than 20Hz are averaged out; 
(c) P5abc$\Delta$A; there is a slow switching between the
closed molecule and the molecule with $H1$ opened and $H2$ 
opening or closing on a shorter time scale,
 the  opening of $H3$  (distance between 
extremities of 31 nm) is a rare event at the force of 12.9 pN;  
(d) the hypothetical RNA molecule of Fig. 1d.}
\label{fig2}
\end{figure}

\begin{center}
\begin{figure}
\includegraphics[height=250pt,angle=-90] {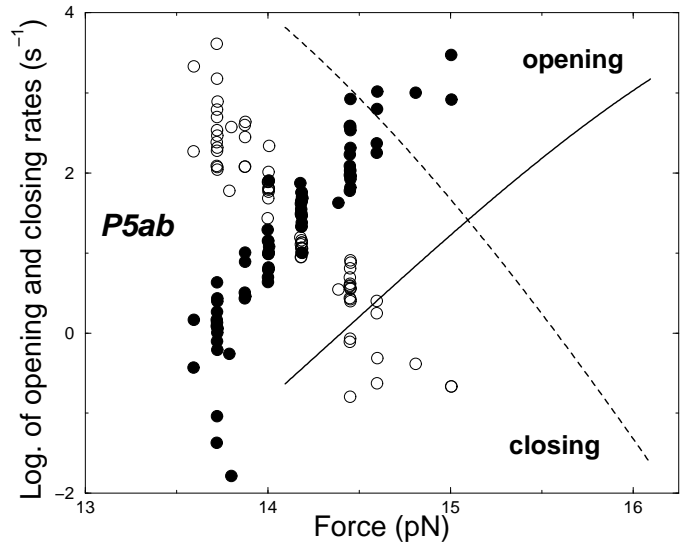}
\vskip 1cm
\caption
{Log. of the opening and closing rates for P5ab as measured in [1]
as a function of force (full circles: opening, empty circles:
closing), compared to theory (full line: opening,
dashed line: closing). The slopes of $\ln k_o, \ln k_c$  
give the relative positions $n_o=8,n_c=11$ 
of the transition state from the closed ($n=3$) and open ($n=24$) 
states respectively, with an absolute location in $n^*\simeq 12$.}
\label{figure4}
\end{figure}
\end{center}


\begin{thebibliography}{99}
\bibitem{noterandom}
Additional contributions to $G^*$ are present for long, random
sequences \cite{Lub}.
For uncorrelated base pair distributions of the free energies
$g_0(i)$, the free energy $G(n,f)$ (2) behaves as a random walk, resulting 
in a maximal barrier $G^* = O(\sqrt N)$.
\bibitem{notep5abcDa}
P5abc$\Delta$A includes three helices with $N_1= 12$ 
(H1), $N_2=5$ (H2), and $N_3=9$ (H3) base pairs. 
A single (respectively double) boundary between open and
closed regions arises if less (resp. more) than 12 base pairs are open.
The number of configurations of the molecule is thus
$N=N_1 +(N_2+1)(N_3+1)=72$. From configurations where H2 and H3 
are partially opened, four transition steps are possible.
\end{thebibliography}
\end{document}